\documentclass[fleqn,10pt]{wlscirep}
\usepackage[utf8]{inputenc}
\usepackage[T1]{fontenc}
\usepackage{xcolor}
\usepackage[left]{lineno}

\title{An experimental study of the surface formation of methane in interstellar molecular clouds}

\author[1,*]{D. Qasim}
\author[1]{G. Fedoseev}
\author[2]{K.-J. Chuang}
\author[1]{J. He}
\author[3]{S. Ioppolo}
\author[4]{E. F. van Dishoeck}

\author[1]{H.~Linnartz}
\affil[1]{Laboratory for Astrophysics, Leiden Observatory, Leiden University, PO Box 9513, NL--2300 RA Leiden, The Netherlands}
\affil[2]{Laboratory Astrophysics Group of the Max Planck Institute for Astronomy at the Friedrich Schiller University Jena, Institute of Solid State Physics, Helmholtzweg 3, D-07743 Jena, Germany}
\affil[3]{School of Electronic Engineering and Computer Science, Queen Mary University of London, Mile End Road, London E1 4NS, UK}
\affil[4]{Leiden Observatory, Leiden University, PO Box 9513, NL--2300 RA Leiden, The Netherlands}

\affil[*]{dqasim@strw.leidenuniv.nl}

\begin{abstract}

Methane is one of the simplest stable molecules that is both abundant and widely distributed across space. It is thought to have partial origin from interstellar molecular clouds, which are near the beginning of the star formation cycle. Observational surveys of CH$_4$ ice towards low- and high-mass young stellar objects showed that much of the CH$_4$ is expected to be formed by the hydrogenation of C on dust grains, and that CH$_4$ ice is strongly correlated with solid H$_2$O. Yet, this has not been investigated under controlled laboratory conditions, as carbon-atom chemistry of interstellar ice analogues has not been experimentally realized. In this study, we successfully demonstrate with a C-atom beam implemented in an ultrahigh vacuum apparatus the formation of CH$_4$ ice in two separate co-deposition experiments: C + H on a 10 K surface to mimic CH$_4$ formation right before H$_2$O ice is formed on the dust grain, and C + H + H$_2$O on a 10 K surface to mimic CH$_4$ formed simultaneously with H$_2$O ice. We confirm that CH$_4$ can be formed by the reaction of atomic C and H, and that the CH$_4$ formation rate is 2 times greater when CH$_4$ is formed within a H$_2$O-rich ice. This is in agreement with the observational finding that interstellar CH$_4$ and H$_2$O form together in the polar ice phase, i.e., when C- and H-atoms simultaneously accrete with O-atoms on dust grains. For the first time, the conditions that lead to interstellar CH$_4$ (and CD$_4$) ice formation are reported, and can be incorporated into astrochemical models to further constrain CH$_4$ chemistry in the interstellar medium and in other regions where CH$_4$ is inherited.  

\end{abstract}

\begin{document}

\flushbottom
\maketitle
%
%
\thispagestyle{empty}

\section*{Introduction}

 Interstellar methane (CH$_4$) ice has been detected towards low- and high-mass young stellar objects (YSOs), where it has an abundance relative to H$_2$O ice of 1-11\% \cite{boogert2015observations}, and is observationally constrained to be formed primarily from the reaction of H-atoms and solid C at the onset of the H$_2$O-rich ice phase of molecular clouds \cite{oberg2008c2d}. This pathway to CH$_4$ ice formation is also accounted for in astrochemical models to an extent \cite{aikawa2008molecular}. Such a constraint is complemented by the fact that the sequential solid-state reactions, \textrm{C + H} $\rightarrow$ \textrm{CH}, \textrm{CH + H} $\rightarrow$ \textrm{CH$_2$}, \textrm{CH$_2$ + H} $\rightarrow$ \textrm{CH$_3$}, \textrm{CH$_3$ + H} $\rightarrow$ \textrm{CH$_4$}, are likely to be barrierless \cite{cuppen2017grain} and exothermic \cite{nuth2006chemical}, whereas the gas-phase CH$_4$ formation pathway includes rate-limiting steps \cite{smith1989effects}. CH$_4$ has been detected on comets \cite{mumma1996detection,gibb2003methane}, which are thought to have delivered interstellar CH$_4$ to planetary bodies, particularly during the early phases of our Solar System. Indeed, CH$_4$ has been detected in a number of planetary systems \cite{formisano2004detection,swain2008presence,stern2015pluto}, where its origin may be from the interstellar medium (ISM), as reported for the CH$_4$ found in Titan’s atmosphere \cite{mousis2002d, mousis2009primordial}. In essence, CH$_4$ is a ubiquitous species within the star formation cycle, with partial origin from the ISM.

To date, the solid-state formation of CH$_4$ by atomic C and H under conditions relevant to the H$_2$O-rich ice phase in interstellar clouds has not been confirmed in the laboratory, which causes ambiguity to the idea of the origin and interstellar formation of CH$_4$. Experimental investigations have been limited to the hydrogenation of graphite surfaces \cite{bar1980interstellar} and H$_2$O-poor conditions \cite{hiraoka1998gas}, where the reported formation pathways of CH$_4$ are still under debate. The work by Hiraoka et al.\cite{hiraoka1998gas}, which the experimental conditions are closest to the study presented here, consisted of plasma-activated CO gas as the atomic carbon source and did not include a H$_2$O ice matrix. The present work shows CH$_4$ formation starting directly from atomic C in a H$_2$O-rich ice, which is a more realistic interstellar scenario. The lack of experimental evidence on this topic is due to the technical challenges that are associated with the coupling of an atomic carbon source with an ultrahigh vacuum (UHV) setup that is designed to study atom-induced surface reactions under molecular cloud conditions. In this study, we investigate two interstellar relevant reactions for CH$_4$ ice formation: the simultaneous deposition of C + H and C + H + H$_2$O on a 10 K surface. This first experimental investigation of CH$_4$ ice by the reaction of C- and H-atoms under conditions mimicking those of interstellar molecular cloud environments is essential to understanding the distribution of CH$_4$ in the star formation cycle. Such work also provides formation yields, rates, temperature and reactant dependencies -- values which were not previously available in the literature.

\section*{Results}

\begin{figure}[htb!]
\centering
\includegraphics[totalheight=4cm]{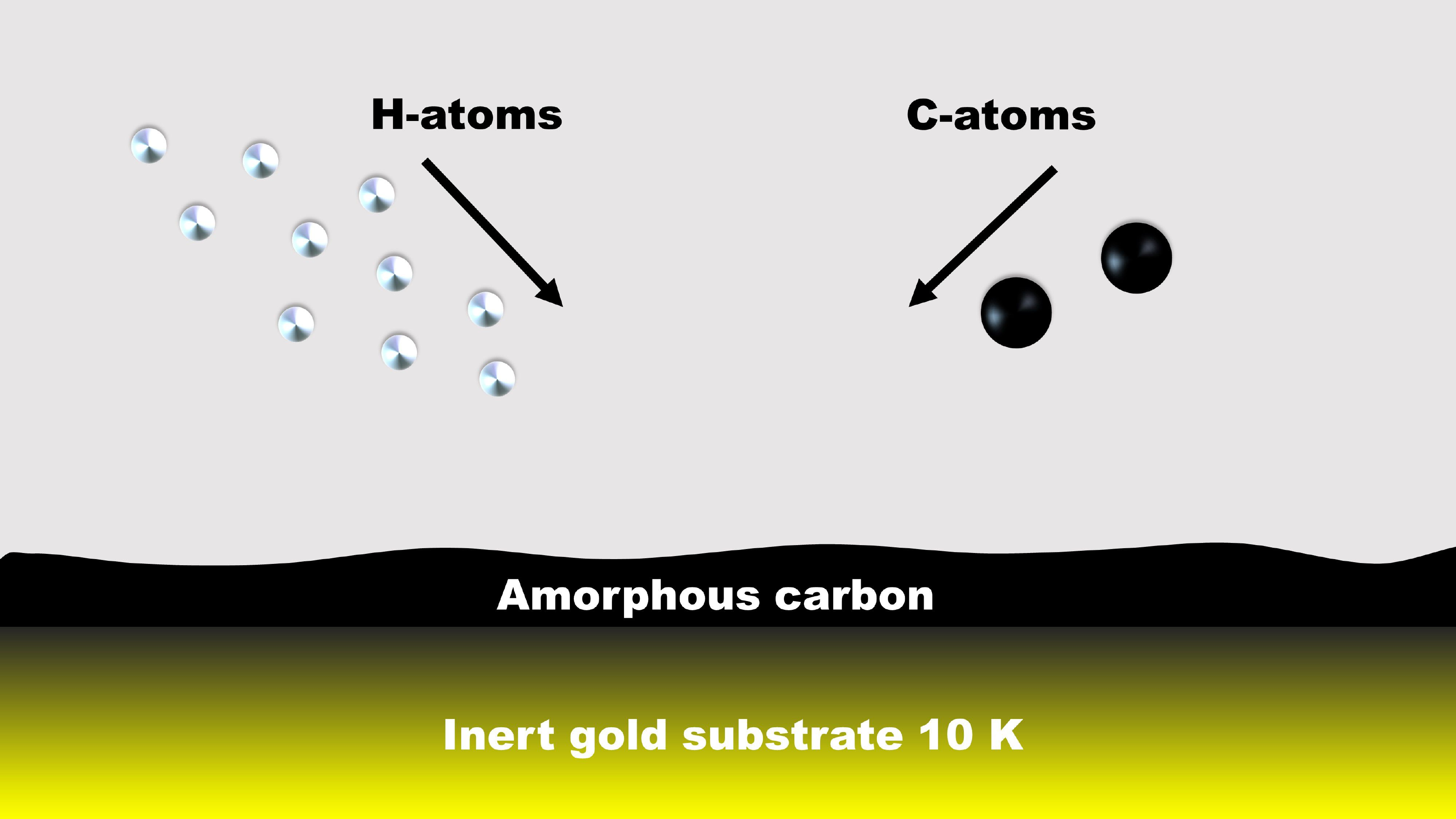}
\includegraphics[totalheight=4cm]{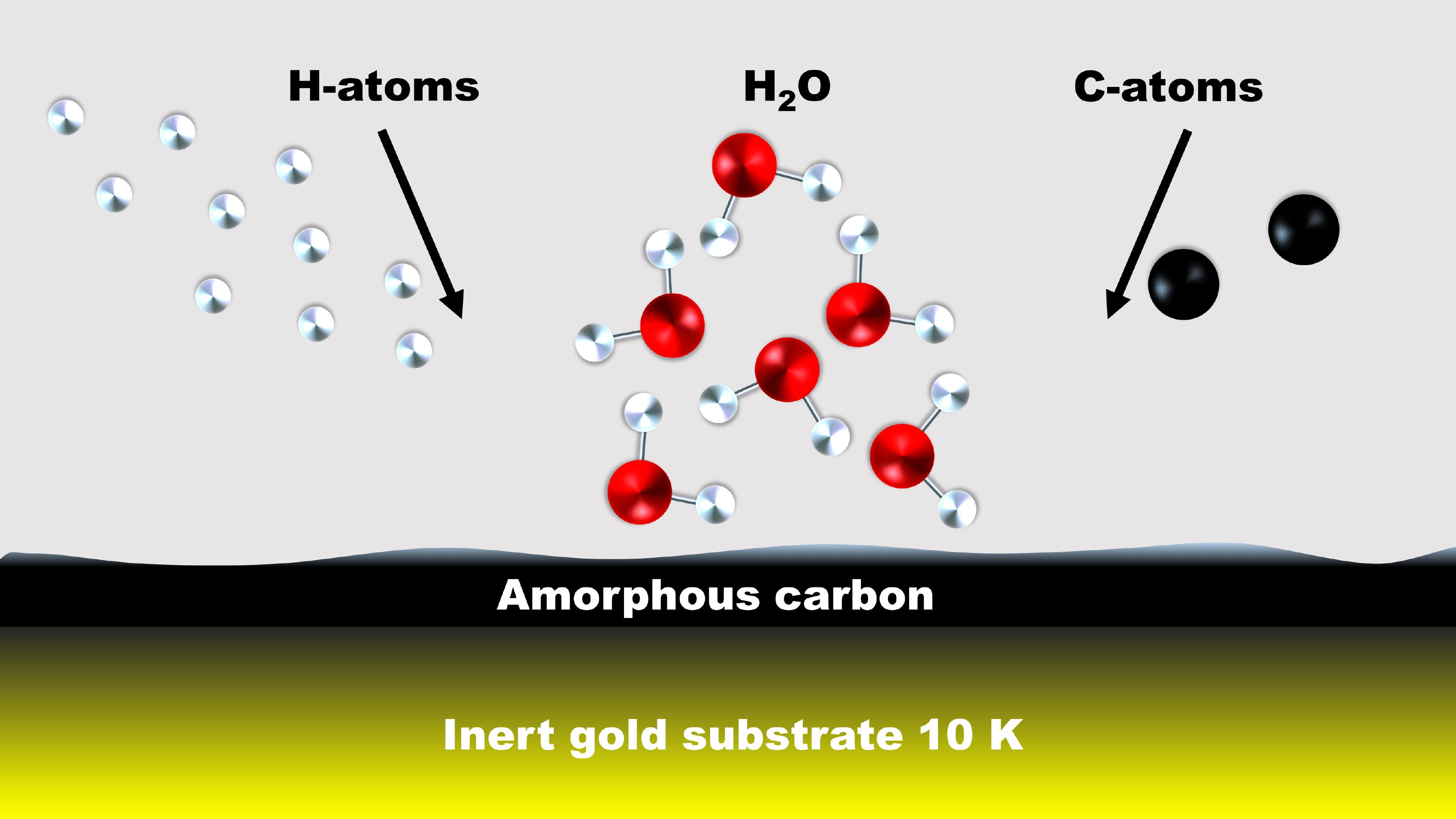}
\caption{Visualization of the two experiments highlighted in this study. (Left) The simultaneous deposition of C- and H-atoms on a 10 K carbonaceous surface is shown. (Right) The addition of H$_2$O molecules is illustrated. Note that the formation of carbonaceous layers is due to the high sticking of C-atoms and available flux. The angles of deposition are arbitrarily displayed.}
\label{fig1}
\end{figure}

Figure~\ref{fig1} provides a visual of the two investigated experiments, whereas more details are found in the Methods section. Product identification is unambiguously shown by the \textit{in situ} detection technique, reflection absorption infrared spectroscopy (RAIRS). A list of the experiments performed and the following formation yields are provided in Table~\ref{table30.1}. The flux of H$_2$O ($\sim$$6 \times 10^{12}$ molecules cm$^{-2}$ s$^{-1}$) was chosen to create a CH$_4$:H$_2$O ratio of 10\% for experiments 2.1-2.3 in order to reflect the composition of interstellar ices \cite{boogert2015observations}. It is important to note that H$_2$O is simultaneously deposited with C and H to create a mixed CH$_4$ and H$_2$O ice, as would be found in cold interstellar clouds, and is not meant to represent the accretion of gas-phase H$_2$O in such environments. The flux of H-atoms used was $\sim$$9 \times 10^{12}$ atoms cm$^{-2}$ s$^{-1}$, in comparison to $\sim$$5 \times 10^{11}$ atoms cm$^{-2}$ s$^{-1}$ for that of C-atoms, which is representative of the dominance of H-atoms over C-atoms in the ISM. Each experimental set has the purpose of disentangling other possible CH$_4$ formation routes. Additionally, the table provides information on how various experimental conditions influence the formation of CH$_4$ (CD$_4$) when H$_2$O is present and/or absent. The results from each experiment are discussed below.

\begin{table*}[htb!]
	\centering
	\caption{A list of experiments, along with the experimental parameters and subsequent product yields. Note that experiments 1.1-1.3 represent the same experiment, but with varying fluences (also with experiments 2.1-2.3). (-) and (<) refer to not applicable and non-detections, respectively. Details on band strength determination for column density calculations are found in the Methods section. The reported CH$_4$ column densities are overestimated by $<$ 25\%, as C can possibly react with H$_2$/D$_2$ in the H$_2$O/D$_2$O experiments to form CH$_4$/CD$_4$, but not with H$_2$O/D$_2$O, as further discussed in the Supporting Information (see Figures~\ref{fig4} and~\ref{fig6}).}
	\label{table30.1}
	\resizebox{\textwidth}{!}{
	\begin{tabular}{c c c c c c c} 
		\hline
		No. & Experiments & T$_{\mathrm{sample}}$ & Column density$_{{\mathrm{{{CH}_4}/{CD}_4}}}$ & Column density$_{\mathrm{{{H}_2}O}}$ & Ratio$_{\mathrm{{CH_4}:{H_2O}}}$ & Time\\
& & (K) & (molecules cm$^{-2}$) & (molecules cm$^{-2}$) & (\%) & (s) \\ 
		\hline
		
        1.1 & C + H &  10 & $2.8 \times 10^{14}$ & - & - & 1440\\ 
        1.2 & C + H &  10 & $2.5 \times 10^{14}$ & - & - &1080\\ 
        1.3 & C + H &  10 & $2.1 \times 10^{14}$ & - & - &720\\ 
       
        2.1 & C + H + H$_{2}$O & 10 &  $8.1 \times 10^{14}$ & $8.0 \times 10^{15}$ & 10 &1440\\ 
        2.2 & C + H + H$_{2}$O & 10 &  $6.4 \times 10^{14}$ & $6.4 \times 10^{15}$ & 10 &1080\\ 
        2.3 & C + H + H$_{2}$O & 10 &  $4.3 \times 10^{14}$ & $4.2 \times 10^{15}$ & 10 &720\\ 
        2.4 & C + H$_2$ + H$_{2}$O & 10 &  $2.0 \times 10^{14}$ & $4.1 \times 10^{15}$ & 5 &1440\\ 
        3 & C + D + H$_{2}$O & 10 &  $7.7 \times 10^{14}$* & $7.6 \times 10^{15}$ & 10 &1440\\ 
        4 & C + H + H$_{2}$O &  25 & < $4.2 \times 10^{13}$ & $7.2 \times 10^{15}$ & < 0.6 &1440\\ 

		\hline
	\end{tabular}}
	\begin{tablenotes}
	*Cannot directly compare to CH$_4$ column densities. See main text for more details.\\
	\end{tablenotes}
    \end{table*}

\begin{figure}[htb!]
\centering
\includegraphics[totalheight=6cm]{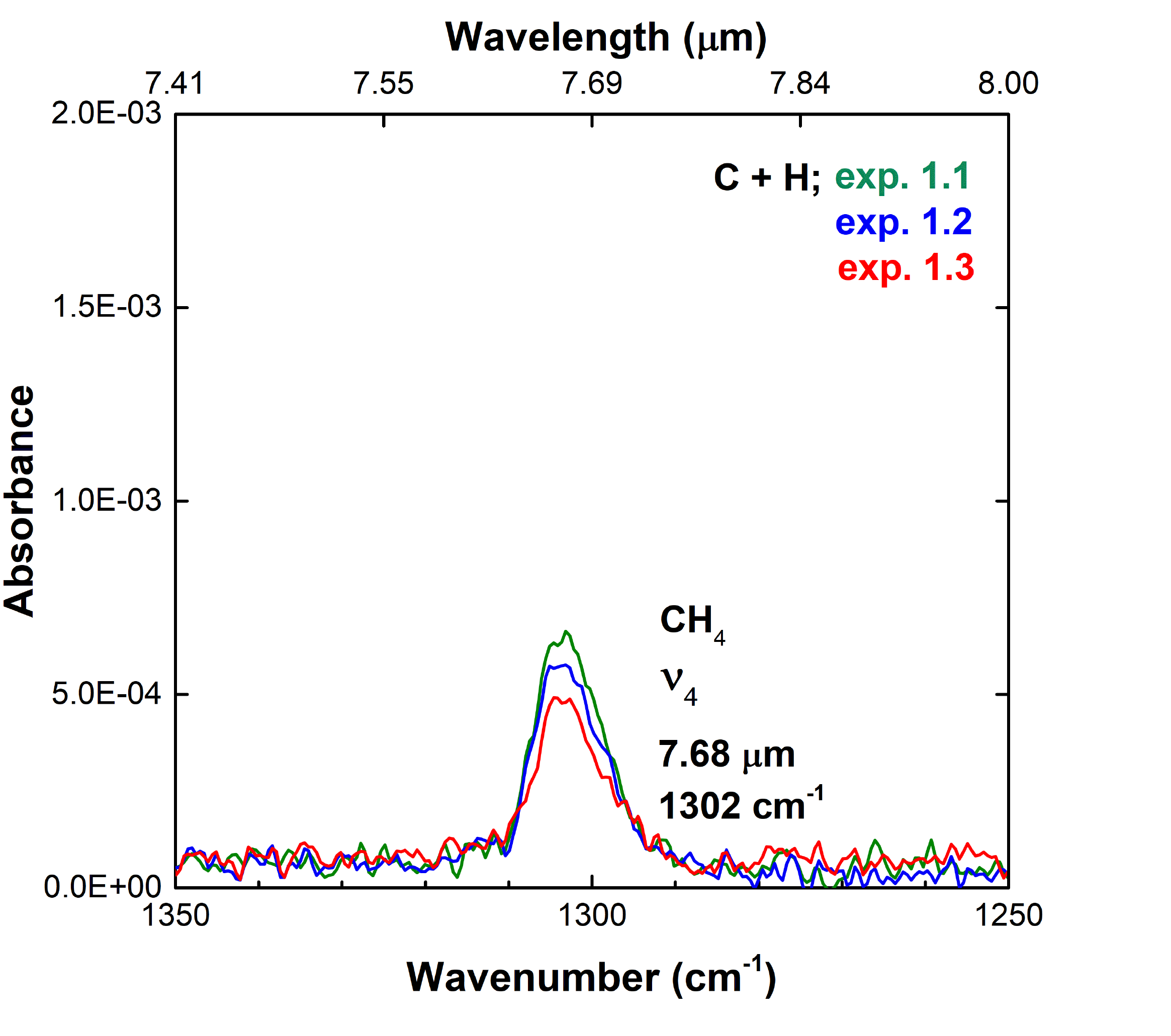}
\includegraphics[totalheight=6cm]{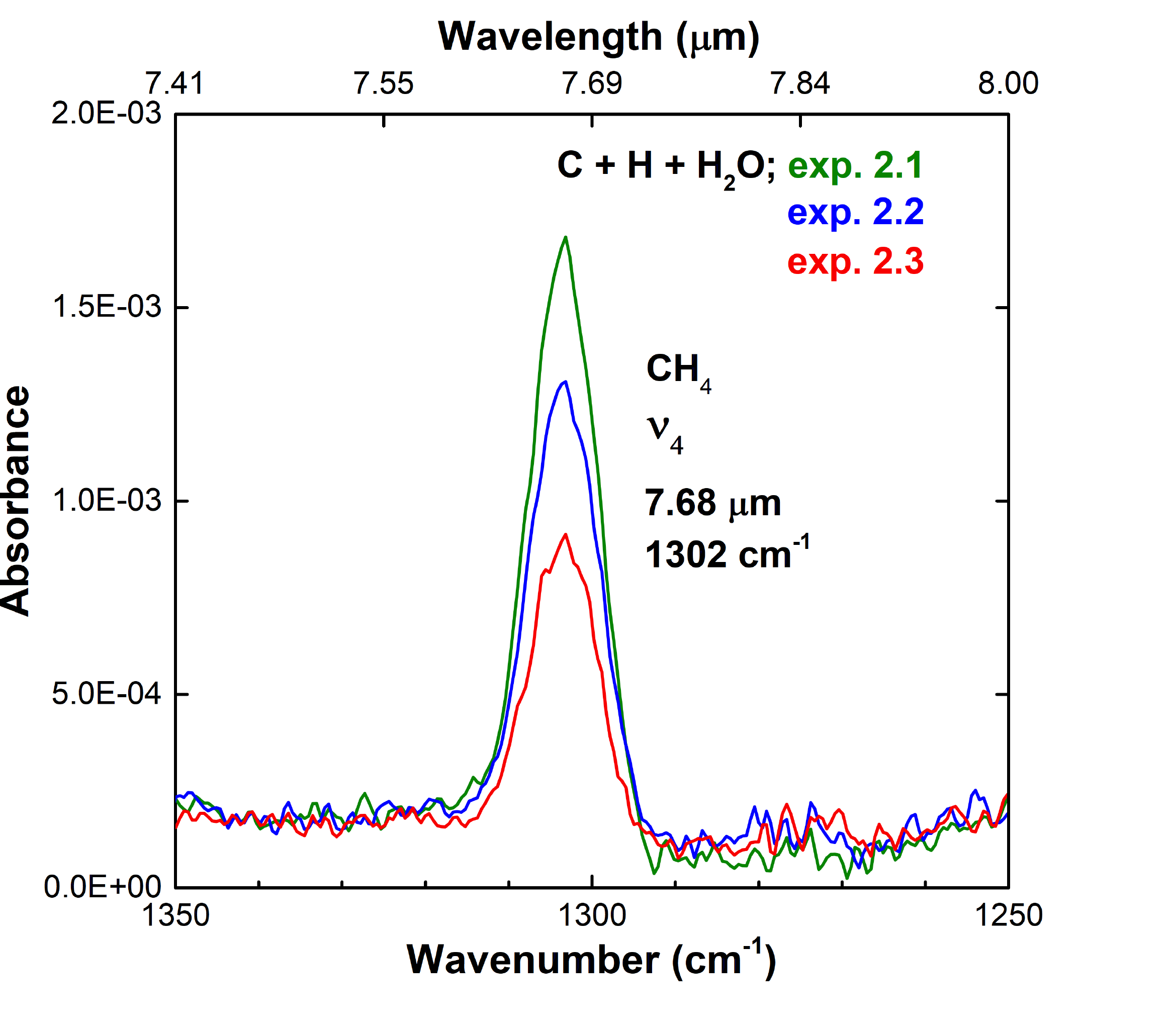}
\caption{(Left) RAIR spectra, in which only the selected feature of interest is shown (i.e., CH$_4$ $\nu$$_4$ mode), were acquired after co-deposition of C + H on a 10 K surface in 360 second intervals (exps. 1.1-1.3), and (right) after co-deposition of C + H + H$_2$O on a 10 K surface in 360 second intervals (exps. 2.1-2.3). RAIR spectra are offset for clarity.}
\label{fig2}
\end{figure}

The RAIR spectra reflecting the experiments of 1.1-1.3 and 2.1-2.3 in Table~\ref{table30.1} are displayed in Figure~\ref{fig2}. The left and right panels unambiguously confirm CH$_4$ formation by featuring the very strong $\nu$$_4$ mode of CH$_4$ \cite{chapados1972infrared} at various deposition times. Extra confirmation of CH$_4$ formation by D-isotope substitution and appearance of the $\nu$$_3$ mode are provided in the Supporting Information, Figure~\ref{fig5}. In both, C + H and C + H + H$_2$O experimental sets, CH$_4$ formation is observed within minutes, in addition to no detection of CH$_{n}$ radicals or their recombination products such as C$_2$H$_2$, C$_2$H$_4$, and C$_2$H$_6$. This is consistent with the efficient recombinations between CH$_{n}$ radicals and H-atoms, and that the lifetime of such radicals is relatively short under our experimental conditions. This also implies that the competing H-abstraction reactions do not dominate in either case. The abstraction reactions, \textrm{CH + H} $\rightarrow$ \textrm{C + H$_2$} and \textrm{CH$_2$ + H} $\rightarrow$ \textrm{CH + H$_2$}, are reported to be essentially barrierless \cite{baulch1992evaluated,harding1993theoretical}, whereas the barrier for \textrm{CH$_3$ + H} $\rightarrow$ \textrm{CH$_2$ + H$_2$} is reported to have a high value of $\sim$7600-7700 K \cite{baulch1992evaluated}. \textrm{CH$_4$ + H} $\rightarrow$ \textrm{CH$_3$ + H$_2$} also has a high barrier height of $\sim$7450-7750 K \cite{corchado2009hydrogen}, and thus may explain why CH$_4$ continues to form despite some abstraction reactions competing with addition reactions.

It is apparent from Table~\ref{table30.1} and the panels of Figure~\ref{fig2} that CH$_4$ formation is more efficient in the C + H + H$_2$O experiment. The formation rate of CH$_4$ in the C + H experiment is no greater than $3.5 \times 10^{11}$ molecules cm$^{-2}$ s$^{-1}$, and in the C + H + H$_2$O experiment, it is $5.6 \times 10^{11}$ molecules cm$^{-2}$ s$^{-1}$. Note that the total C + H kinetic curve (until CH$_4$ saturation) is non-linear. After 1440 seconds of deposition, the total column density of CH$_4$ amounts to $2.8 \times 10^{14}$ molecules cm$^{-2}$ and $8.1 \times 10^{14}$ molecules cm$^{-2}$ for the C + H and C + H + H$_2$O experiments, respectively. Thus, despite the H$_2$O-ice matrix, which could potentially block C- and H-atoms from meeting each other, formation of CH$_4$ is actually enhanced when H$_2$O is included. This phenomenon is likely, in part, due to the increase in the sticking coefficient of atomic H when H is in the presence of amorphous solid water (ASW) \cite{oberg2010effect,veeraghattam2014sticking}, and the increase in the surface area that atoms can stick to \cite{mayer1986astrophysical}. The sticking probability of H-atoms with an incident energy of 300 K on a 10 K ASW ice is 0.4 \cite{veeraghattam2014sticking}, in comparison to the lower value of $\sim$0.02 for that of graphite \cite{lepetit2011sticking}. Note that the carbon allotrope formed from the atomic source is determined to be amorphous \cite{albar2017atomic}, and thus the sticking coefficient of H on our carbon surface is likely higher than $\sim$0.02.   

The surface formation mechanism is probed in the C + H + H$_2$O experiment at 25 K (exp. 4, Figure~\ref{fig66.5} of the Supporting Information). The formation of CH$_4$ at a deposition temperature of 25 K is negligible in comparison to the formation of CH$_4$ in exp. 2.1, and additionally the CH$_4$ feature is within the level of the noise. This finding indicates that the temperature of the surface largely influences the reaction probability, as the residence time of H-atoms drastically drops above $\sim$15 K \cite{fuchs2009hydrogenation,ioppolo2010water,chuang2015h}. Thus, the Langmuir-Hinshelwood mechanism, which is temperature dependent, is predominant in the formation of CH$_4$ at 10 K.

\section*{Astrochemical Implications and Conclusions}

With the utilization of a UHV setup designed, in part, to investigate the simultaneous accretion of C- and H-atoms in two interstellar relevant ices, we experimentally confirm that CH$_4$ formation proceeds and is more favored when H$_2$O is added to the C + H reaction at 10 K. This supports the conclusions of the observational survey of CH$_4$ ice by {\"O}berg et al. (2008)\cite{oberg2008c2d} that much of the detected CH$_4$ is found in the polar phase of ice evolution and formed by atomic C and H. The expected main route to CH$_4$ formation in our experiments is H-atom addition to C-atoms in a H$_2$O ice:\\

$\textrm{C + H}  \underset{\text{H$_2$O}}{\to} \textrm{CH}, \textrm{CH + H} \underset{\text{H$_2$O}}{\to} \textrm{CH$_2$},
\textrm{CH$_2$ + H} \underset{\text{H$_2$O}}{\to} \textrm{CH$_3$},
\textrm{CH$_3$ + H}  \underset{\text{H$_2$O}}{\to} \textrm{CH$_4$}$

which predominantly follows a Langmuir-Hinshelwood mechanism at 10 K.

The findings presented here parallel the assumption used in models that the sequential hydrogenation of C is barrierless. It is suggested that astrochemical models take into account that the formation of interstellar CH$_4$ should still proceed when C and H accrete onto the growing polar ice, as it is experimentally shown that H$_2$O enhances the probability for C and H to react. Whether the CH$_4$ abundance will substantially increase due to the presence of H$_2$O on an interstellar dust grain needs to be tested in a model. As the astronomically observed CH$_4$-H$_2$O correlation towards YSOs is predominantly set by the availability of the simultaneous accretion of C, O, and H, the CH$_4$ formation rate factor of 2 in a H$_2$O-rich ice experiment is not expected to directly lead to a CH$_4$ formation rate factor of 2 in an astrochemical model. Additionally, in an interstellar ice, C will compete with other species to react with H. Thus, it is also suggested that the assumed rate of $2 \times 10^{11}$ s$^{-1}$ used in models for barrierless reactions (\emph{private communication, H. Cuppen}) should be multiplied by 1 < x < 2 for CH$_4$ formation in a H$_2$O-rich ice by the sequential hydrogenation of C.

This work shows that CH$_4$ can be formed in the solid-state under conditions relevant to interstellar clouds, without the need for extra heat or `energetic' particles. On the other hand, UV photons (or enhanced cosmic rays) are needed to maintain a high abundance of atomic C and O in the gas-phase, which can accrete onto grains to make CH$_4$ and H$_2$O ices. Thus, the early low density translucent cloud phase is optimally suited to make both ices simultaneously and abundantly. In the later denser cloud phases, most gaseous carbon has been transformed into CO, which -- after freeze-out onto the grains -- can be transformed into complex organic ices, and, under cold protoplanetary disk conditions, ultimately to CH$_4$ and hydrocarbon ices \cite{bosman2018co}. It is clear that the reaction proceeds effectively at lower versus higher temperatures, and is enhanced in a H$_2$O matrix, both of which are in-line with astronomical observations. Astrochemical modeling is necessary to take into account the available C-atom fluxes in the ISM in order to place the present findings into a cosmochemical picture. Such dedicated models can then be used to aid in the interpretation of CH$_4$ ice observations with the upcoming James Webb Space Telescope (JWST), as CH$_4$ is best observed with space-based observatories. The increase in sensitivity of the JWST Mid-Infrared Instrument is expected to allow observations of CH$_4$ ice towards numerous background stars to probe more quiescent environments, in addition to observations towards YSOs. The work presented here is the first experimental proof of solid-state CH$_4$ formed in a polar ice. The incorporation of a pure C-atom channel in interstellar networks is as important as including the N-atom channel for the formation of NH$_3$ \cite{fedoseev2014low} and the O-atom channel for the formation of H$_2$O \cite{ioppolo2008laboratory}. As molecular clouds are the universal starting point in the star and planet formation process, this also means that much of the chemical inventory in protoplanetary disks and possibly planets is due to the chemical processes that take place on icy dust grains in molecular clouds prior to collapse. 

The present approach also has applications beyond the formation of CH$_4$. It becomes possible to focus on COM formation through C-atom addition \cite{charnley1997astronomical,giovannelli2001bridge,charnley2005pathways,charnley2009theoretical}. The carbon backbone of COMs has already been proven to be formed by the reaction between C-bearing radicals, such as between HCO and CH$_2$OH \cite{chuang2015h}. This new experimental way of forming COMs by adding atomic C will aid in better understanding the origin of detected COMs, as astrochemical models will be able to take into account the relevance of C-atom addition reactions by including data from experimental simulations, such as those found here.

\section*{Methods}
\label{methods}

The experiments presented in this article were performed with SURFRESIDE$^{2}$, which is a UHV apparatus that allows qualitative and quantitative investigations of the ice chemistry of molecular clouds. The initial design is discussed in the work by Ioppolo et al. (2013)\cite{ioppolo2013surfreside2}, and details on the recent modifications are provided by Qasim et al. (2019a)\cite{qasim2019alcohols}. The apparatus partially consists of three atomic beam line chambers that are connected to a main chamber, which reaches a base pressure of $2-3 \times 10^{-10}$ mbar prior to co-deposition. Within the middle of the chamber is a closed-cycle helium cryostat that has a gold-plated copper sample used to grow ices. On top of the sample lies a coating of carbon that is visible to the naked eye, and has been characterized to be amorphous when originating from the atomic carbon source \cite{albar2017atomic}. Due to the high sticking coefficient of atomic carbon, formation of these carbonaceous layers is difficult to avoid. Resistive heating of a cartridge heater was applied to heat the sample. With the incorporation of a sapphire rod, the sample is able to be cooled to a low temperature of 7 K and heated to a high temperature of 450 K. Such temperatures were measured by a silicon diode sensor that has an absolute accuracy of 0.5 K.

In this study, two of the three atomic beam lines were used to create atoms, where the base pressure of both atomic chambers was $2-3 \times 10^{-9}$ mbar. A Microwave Atom Source (MWAS; Oxford Scientific Ltd.), which consists of a 2.45 GHz microwave power supply (Sairem) that was operated at 200 W, was employed to produce H- and D-atoms from H$_2$ and D$_2$, respectively, with a dissociation rate that is less than unity. A nose-shaped quartz tube is attached at the exit of the source so that atoms can depart energy through collisions with the tube before reaching the cooled sample. An atomic carbon sublimation source (SUKO-A 40; Dr. Eberl MBE-Komponenten GmbH), which uses a power supply (Delta Elektronika, SM 15-100) to induce carbon sublimation, was exploited to create ground state C-atoms. Graphite powder was packed within a tantalum tube that was heated to around 2300 K, which leads to a reaction between molecular carbon and tantalum to form TaC$_x$. This process ultimately breaks apart molecular carbon into atomic carbon. Due to the high sticking coefficient of atomic carbon, a quartz pipe was not incorporated to cool the atoms prior to deposition. However, the heat of the C-atoms involved in the initial step of CH$_4$ formation is not expected to qualitatively affect the results, as C + H is expected to be barrierless, and $\textrm{C + H$_2$}$ $\rightarrow$ $\textrm{CH + H}$ is highly endothermic \cite{harding1993theoretical,guadagnini1996unusual}. The third atomic beam line, a Hydrogen Atom Beam Source (HABS) \cite{tschersich1998formation, tschersich2000intensity, tschersich2008design}, was not used as a source of H-atoms in this work.

Details on the preparation of gases and liquids used to create the interstellar ice analogues are described below. H$_2$ (Linde 5.0) and D$_2$ (Sigma-Aldrich 99.96\%) gases were transferred into the MWAS vacuum chamber. A H$_2$O sample was connected to the HABS chamber, where the H$_2$O was cleaned before every experiment by one freeze-pump-thaw cycle. Note that H$_2$O was not formed on the surface but instead deposited, as the focus is to disentangle the formation routes to CH$_4$. All prepared gases and liquids were released into the main chamber by leak valves.

The reflection absorption infrared spectroscopy (RAIRS) technique was performed to probe product formation \textit{in situ}, as well as obtain quantitative information about the products formed through analysis of their vibrational modes. A Fourier Transform Infrared Spectrometer (FTIR), with a fixed scan range of 4000 - 700 cm$^{-1}$ and resolution of 1 cm$^{-1}$, was applied in the RAIRS study. To measure the abundances of CH$_4$/CD$_4$ formed and the CH$_4$/CD$_4$:H$_2$O ice abundance ratios, a modified Lambert-Beer equation was used. Band strength values of $4.40 \times 10^{-17}$ cm molecule$^{-1}$, $2.20 \times 10^{-17}$ cm molecule$^{-1}$, and $4.95 \times 10^{-17}$ cm molecule$^{-1}$ were used to calculate the column densities of CH$_4$ ($\nu$$_4$ mode; 1302 cm$^{-1}$), CD$_4$ ($\nu$$_4$ mode; 993 cm$^{-1}$), and H$_2$O ($\nu$$_2$ mode; 1665 cm$^{-1}$), respectively. The initial values were obtained from Bouilloud et al. (2015)\cite{bouilloud2015bibliographic} for CH$_4$ and H$_2$O, and from Addepalli \& Rao (1976)\cite{addepalli1976infrared} for CD$_4$. A transmission-to-RAIR setup determined proportionality factor of 5.5 was then applied. The proportionality factor was calculated using the band strength of CO (2142 cm$^{-1}$) measured on our setup through the laser interference technique, where the band strength is reported in Chuang et al. (2018)\cite{chuang2018reactive}. 

To secure that the CH$_4$ formation rate is higher in the H$_2$O-rich experiment, the repeatability of the C + H and C + H + H$_2$O experiments was evaluated. As the formation rates are determined by plotting the CH$_4$ column densities as a function of time, the uncertainty in the formation rates and column densities can be assessed by measuring the relative standard deviation (RSD) between data points of the same experiment that was performed on different days. For the C + H and C + H + H$_2$O experiments, average RSD values of 10\% and 2\% were measured, respectively. Such values further secure the claim that the CH$_4$ formation rate is larger in a H$_2$O-rich ice. The higher repeatability of the C + H + H$_2$O experiment is predominantly due to the more accurate column density measurement of CH$_4$, as the S/N of the 1302 cm$^{-1}$ feature is greater.   


\section*{Author Contributions} D.Q. performed the experiments and wrote the manuscript. D.Q. and G.F. designed the experiments and analyzed the data. K.J.C. helped with column density measurements and error calculations. J.H. and S.I. provided insights on the surface formation mechanism. E.F.vD. and H.L. generously assisted with the astrochemical implications. H.L. initiated the project. All authors participated in discussion of the experiments, analysis and interpretation of the results, and shaping the manuscript.

\section*{Acknowledgements}

This research benefited from the financial support from the Dutch Astrochemistry Network II (DANII). Further support includes a VICI grant of NWO (the Netherlands Organization for Scientific Research) and A-ERC grant 291141 CHEMPLAN. Funding by NOVA (the Netherlands Research School for Astronomy) is acknowledged. D.Q. acknowledges Jordy Bouwman and Edith Fayolle for stimulating discussions. S.I. recognises the Royal Society for financial support and the Holland Research School for Molecular Chemistry (HRSMC) for a travel grant.

\begin{suppinfo}

\renewcommand{\thefigure}{S\arabic{figure}}
\renewcommand{\thetable}{S\arabic{table}}
\renewcommand{\thesection}{S\arabic{section}}
\setcounter{section}{0}
\setcounter{figure}{0}
\setcounter{table}{0}

\section*{Supporting Information}
\label{SI}

Figure~\ref{fig4} shows RAIR spectra in the range of 1350 - 1250 cm$^{-1}$ in order to compare the control experiment of C + H$_2$ + H$_2$O with C + H(H$_2$) + H$_2$O, as not all of the H$_2$ is converted into H in the MW source. Thus, atomic C may participate in a sequence of reactions involving H$_2$ to ultimately form CH$_4$. As both experiments were performed under the same parameters (flux, deposition time, and temperature), the abundances of CH$_4$ formed can be compared. The column densities of CH$_4$ in the C + H(H$_2$) + H$_2$O and C + H$_2$ + H$_2$O experiments are $8.1 \times 10^{14}$ molecules cm$^{-2}$ and $2.0 \times 10^{14}$ molecules cm$^{-2}$, respectively. This implies that the upper limit for the C + H$_2$ reaction route contribution towards the total CH$_4$ abundance is 25\%. It is less because the more dominant reaction route, C + H, is omitted in the C + H$_2$ + H$_2$O experiment (i.e., the reaction efficiency is at least a factor of 4 less and likely much more). It is reported that C + H$_2$ barrierlessly leads to CH + H through the intermediate, CH$_2$, although it is endothermic by $\sim$13,450 K \cite{harding1993theoretical,guadagnini1996unusual}, and therefore considered negligible in the presented experiments. Thus, it may be that the minor amount of CH$_4$ formed starting from H$_2$ is at least due to the supposed barrierless reaction of C + H$_2$ to form the intermediate, CH$_2$, followed by two H-atom additions. CH$_2$ is stabilized due to the presence of the surface third body. The barrier for the \textrm{CH$_2$ + H$_2$} $\rightarrow$ \textrm{CH$_3$ + H} reaction is ambiguous \cite{gesser1956photolysis,lu2010shock}. 

\begin{figure}[htb!]
\centering
\includegraphics[totalheight=6cm]{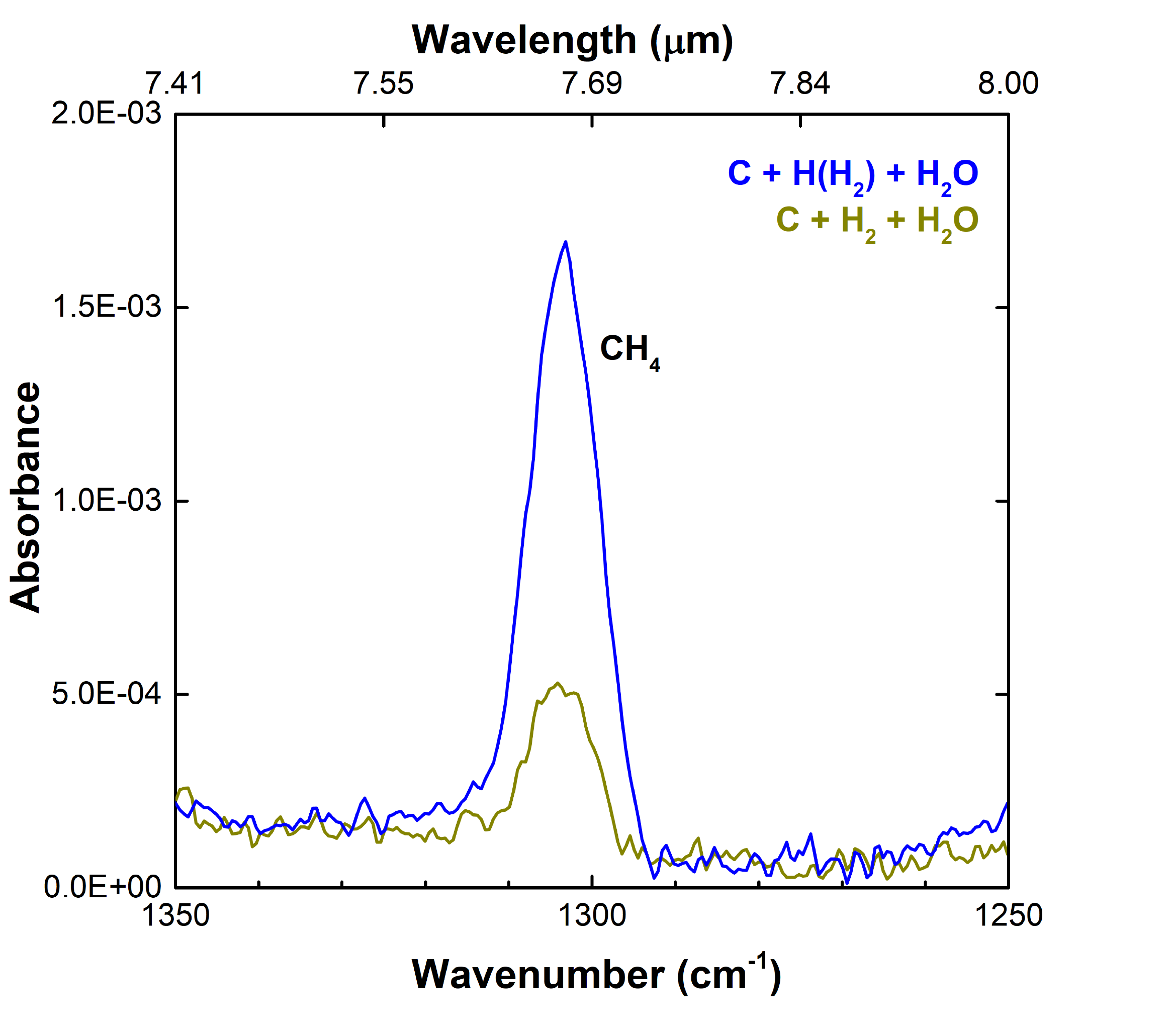}
\caption{RAIR spectrum acquired after co-deposition of C + H(H$_2$) + H$_2$O (exp. 2.1; blue) and C + H$_2$ + H$_2$O (exp. 2.4; green) on a 10 K surface, which shows the minor formation of CH$_4$ starting from H$_2$ versus H. RAIR spectra are offset for clarity.}
\label{fig4}
\end{figure}

Figure~\ref{fig5} provides additional proof for CH$_4$ formation in the C + H + H$_2$O experiments. The left panel shows the isotopic shift of the deformation mode when H-atoms are substituted by D-atoms in the co-deposition experiment. This additionally shows that CD$_4$ can also be formed in a H$_2$O-rich ice via atom-atom reactions, if D-atoms are available for reaction. A CD$_4$ column density of $7.7 \times 10^{14}$ molecules cm$^{-2}$ was measured, which is close to the CH$_4$ column density. However, the abundances cannot be directly compared, as the D-atom flux used was approximately twice less in comparison to that of H-atoms. The right panel shows the strong C-H stretching vibrational mode of CH$_4$. 

\begin{figure}[htb!]
\centering
\includegraphics[totalheight=6cm]{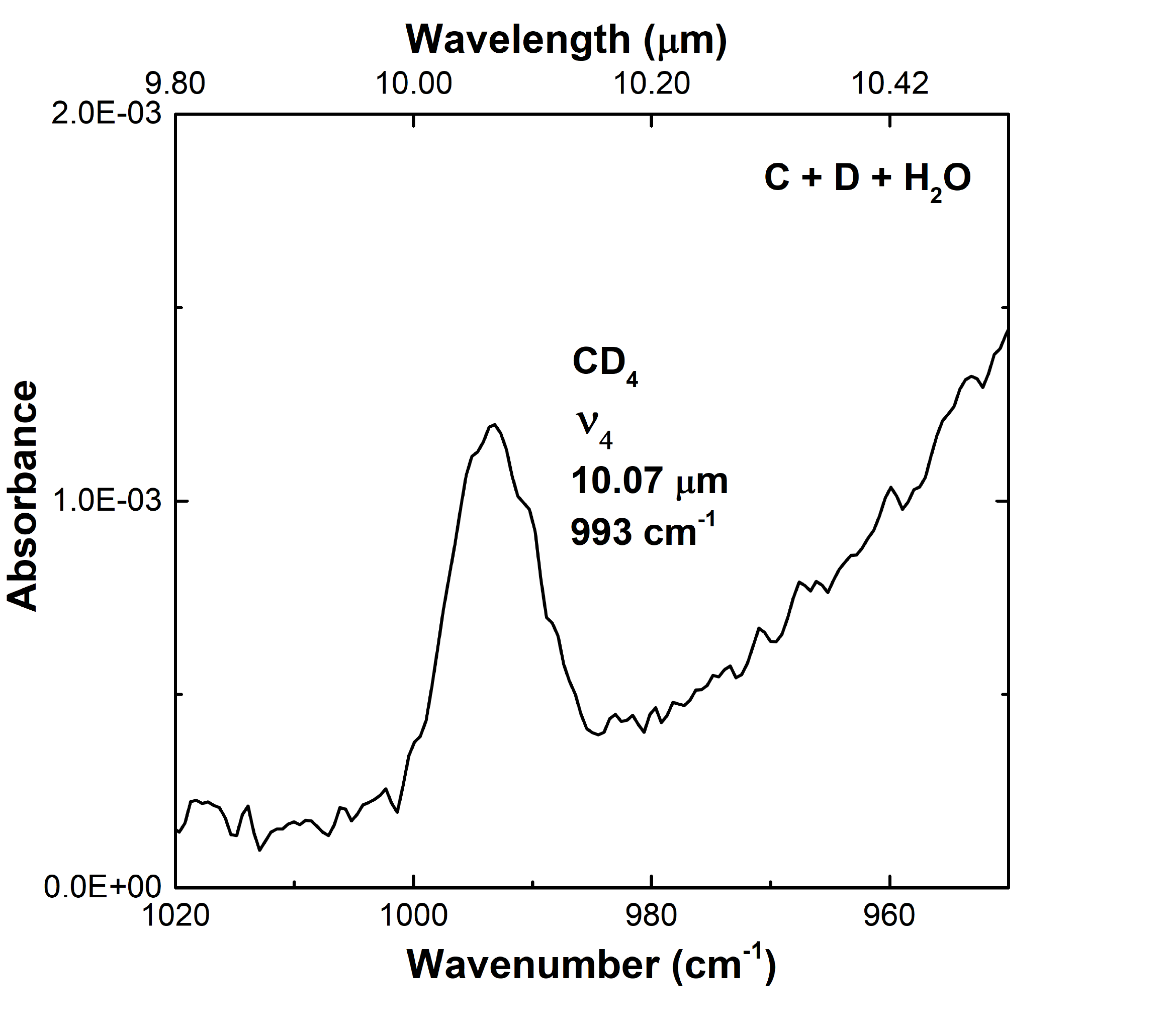}
\includegraphics[totalheight=6cm]{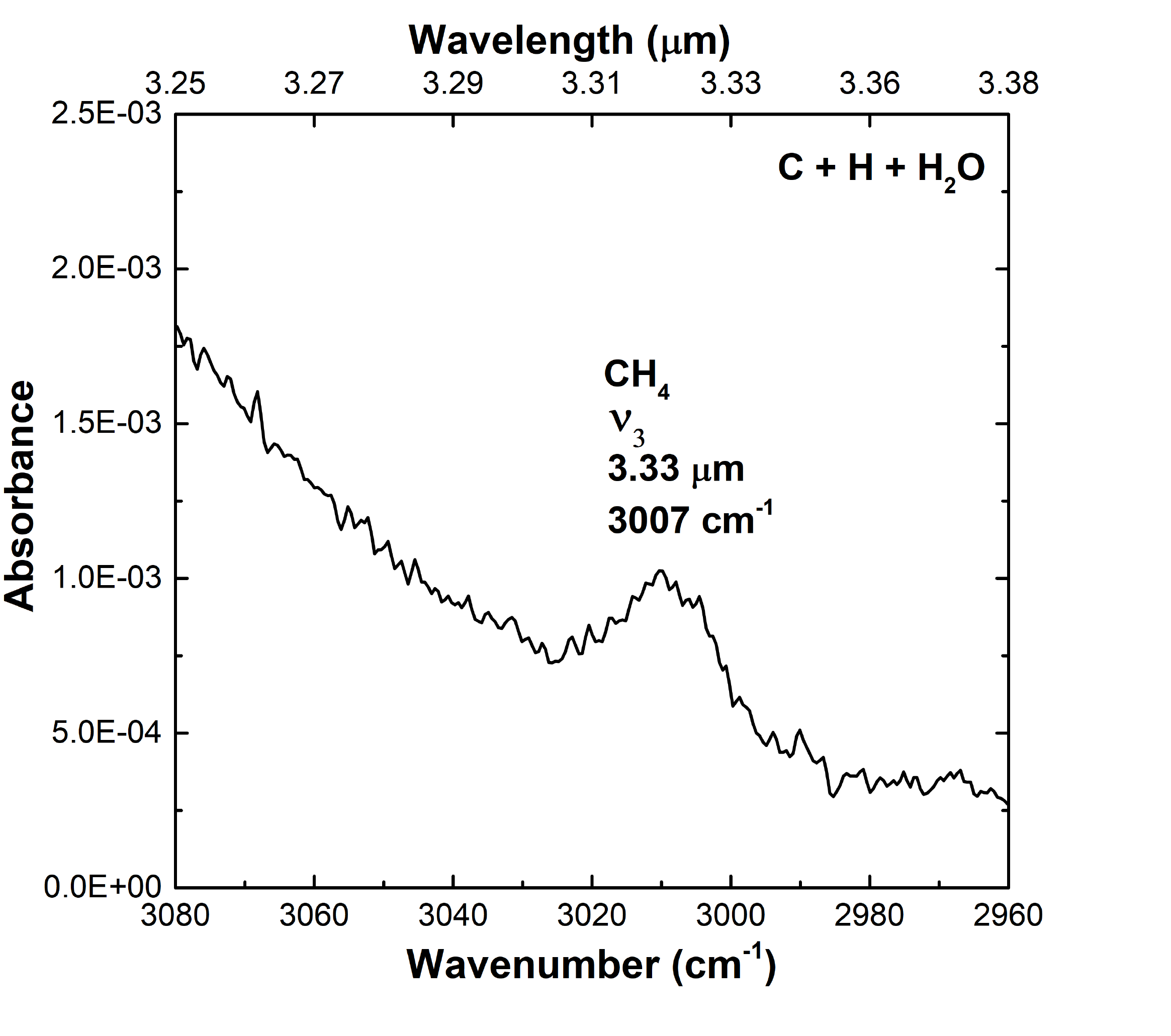}
\caption{The two panels display proof of methane formation. (Left) The $\nu$$_4$ mode in the C + D + H$_2$O experiment (exp. 3), which is intrinsic to that of CD$_4$ \cite{chapados1972infrared}. This is evidence for CH$_4$ formation in the C + H + H$_2$O experiment by observation of the isotopic shift. (Right) Evidence of CH$_4$ formation in the C + H + H$_2$O experiment (exp. 2.1) by observation of the CH$_4$ $\nu$$_3$ mode on the H$_2$O wing. RAIR spectra are offset for clarity.}
\label{fig5}
\end{figure}

The absence of the CH$_4$ signature at $\sim$1300 cm$^{-1}$ in the C + D + H$_2$O experiment (exp. 3) is shown in Figure~\ref{fig6}. This indicates that C and H$_2$O do not react to form CH$_4$ in exp. 3, and therefore should not contribute to form CH$_4$ by abstraction of H-atoms from H$_2$O. This is expected, as the competing reaction of atomic C and H(D) addition is likely barrierless. CD$_3$H, CD$_2$H$_2$, and CDH$_3$ were also not identified. Whether C reacts with the O-atom of H$_2$O will be investigated in a separate dedicated study. Figure~\ref{fig66.5} clearly shows that CH$_4$ formation is negligible at a deposition temperature of 25 K due to the drop in the H-atom residence time on the surface. This extended residence time required for CH$_4$ formation is an indication that both, H-atoms and CH$_n$ intermediates have a period of time available to thermalize with the surface prior to reaction in the 10 K experiments.

\begin{figure}[htb!]
\centering
\includegraphics[totalheight=6cm]{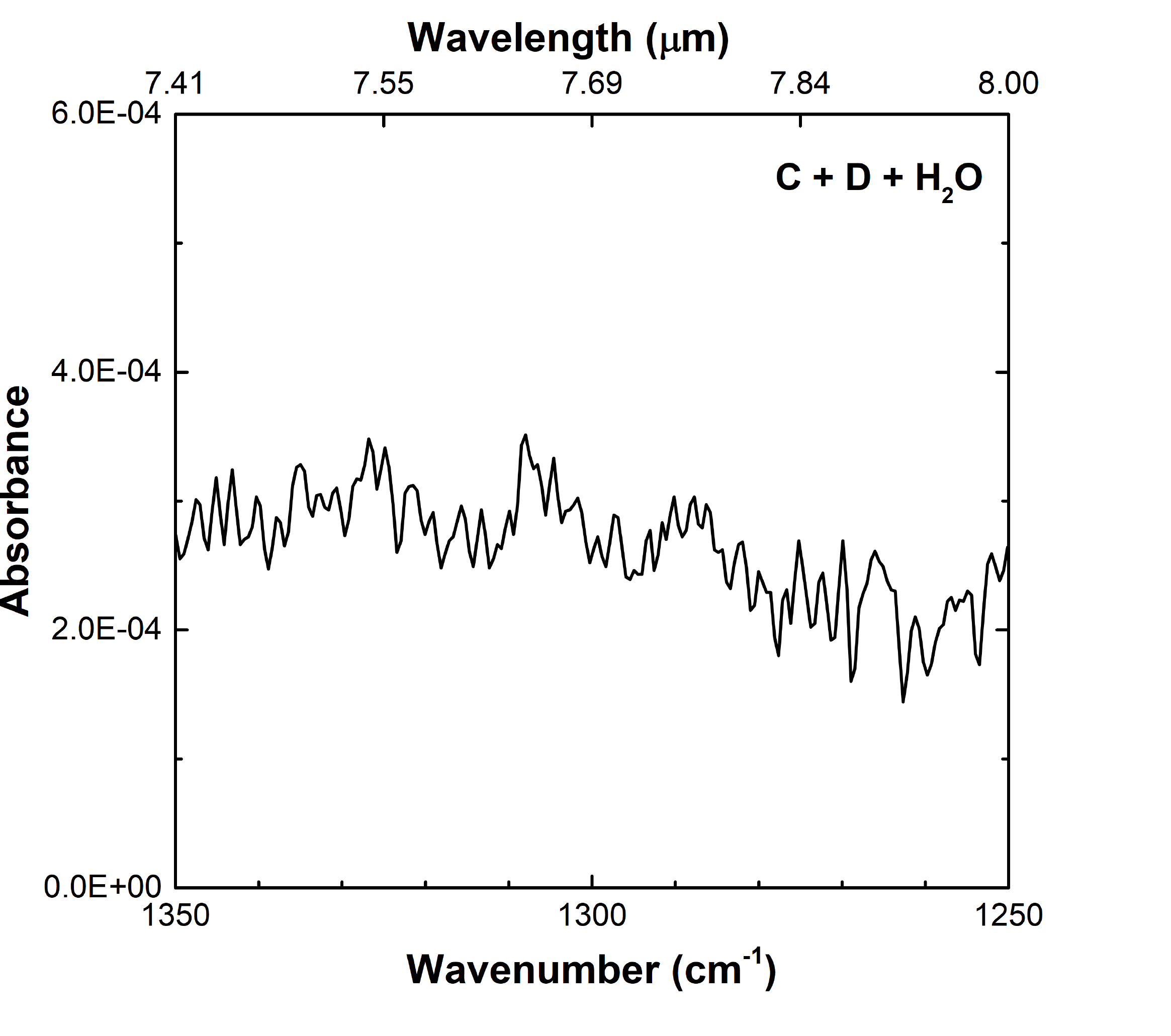}
\caption{Lack of the C-H bending vibrational signal in the C + D + H$_2$O experiment (exp. 3), which shows that C does not react with H$_2$O to form CH$_4$. RAIR spectrum is offset for clarity.}
\label{fig6}
\end{figure}

\begin{figure}[htb!]
\centering
\includegraphics[totalheight=6cm]{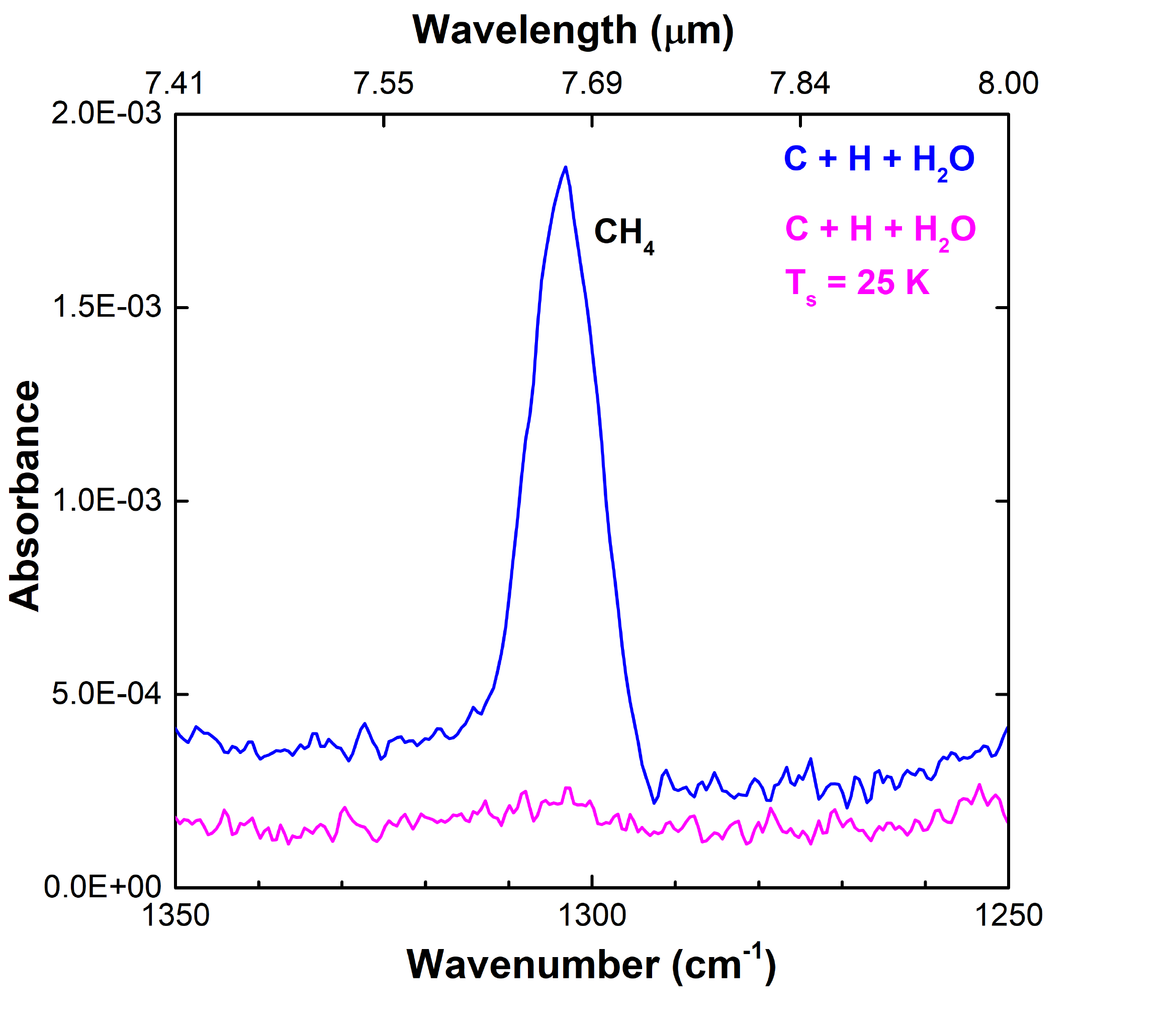}
\caption{RAIR spectra acquired after co-deposition of C + H + H$_2$O on a 10 K surface (exp. 2.1; blue) and co-deposition of C + H + H$_2$O on a 25 K surface (exp. 4; pink), which shows the negligible formation of CH$_4$ at 25 K. RAIR spectra are offset for clarity.}
\label{fig66.5}
\end{figure}

\end{suppinfo}

\clearpage

\end{document}